\documentclass[prd,onecolumn]{revtex4}
\usepackage{dcolumn}
\usepackage{array}
\usepackage{longtable}
\usepackage{multirow}
\usepackage{graphicx}
\usepackage{amssymb}
\usepackage{bm}
\usepackage{hyperref}
\usepackage{epstopdf}
\usepackage{color}
\usepackage{mathrsfs}
\usepackage{amsmath,amssymb,amsthm}
\begin{document}
\title{Determining $H_0$ with the latest HII galaxy measurements}
\author{Deng Wang}
\email{Cstar@mail.nankai.edu.cn}
\affiliation{Theoretical Physics Division, Chern Institute of Mathematics, Nankai University,
Tianjin 300071, China}
\author{Xin-He Meng}
\email{xhm@nankai.edu.cn}
\affiliation{Department of Physics, Nankai University, Tianjin 300071, P.R.China}
\begin{abstract}
We use the latest HII galaxy measurements to determine the value of $H_0$ adopting a combination of model-dependent and model-independent method. By constraining five cosmological models, we find that the obtained values of $H_0$ are more consistent with the recent local measurement by Riess et al. 2016 (hereafter R16) at $1\sigma$ confidence level, and that these five models prefer a higher best-fit value of $H_0$ than R16's result. To check the correctness of $H_0$ values obtained by model-dependent method, for the first time, we implement the model-independent Gaussian processes (GP) using the HII galaxy measurements. We find that the GP reconstructions also prefer a higher value of $H_0$ than R16's result. Therefore, we conclude that the current HII galaxy measurements support a higher cosmic expansion rate.

\end{abstract}
\maketitle
\section{Introduction}
Determining the Hubble constant ($H_0$) accurately is one of the most important challenges in modern cosmology, since it sets the scale for all cosmological times and distances, and it reveals for human beings the present cosmic expansion rate, the size and age of the universe and the cosmic components. Based on the early determination by Hubble \cite{1}, the value of $H_0$ was believed to lie in the large range [50, 100] km s$^{-1}$ Mpc$^{-1}$ \cite{2}. Utilizing improved control of systematics and different calibration techniques, the first precise value, $H_0=72\pm8$ km s$^{-1}$ Mpc$^{-1}$ \cite{3}, was given by the local measurements from the Hubble Space Telescope (HST) in 2001. Ten years later, Riess et al. (hereafter R11) calibrated the Type Ia supernovae (SNe Ia) and obtained $H_0=73.8\pm2.4$ km s$^{-1}$ Mpc$^{-1}$ \cite{4} by using three indicators, i.e., the distance to NGC 4258 from a mega-maser measurement, the trigonometric parallaxes measurements to the Milk Way (MW) Cepheids, Cepheid observations and a modified distance to the Large Magellanic Cloud (LMC). In 2012, there are three groups to measure the Hubble constant: Riess et al. got $H_0=75.4\pm2.9$ km s$^{-1}$ Mpc$^{-1}$ by utilizing the Cepheids in M31 \cite{5}; Freedman et al. obtained $H_0=74.3\pm2.1$ km s$^{-1}$ Mpc$^{-1}$ by using a mid-infrared calibration for the Cepheids \cite{6}. Subsequently, in 2013, $H_0=67.3\pm1.2$ km s$^{-1}$ Mpc$^{-1}$ derived by Planck \cite{7} from the anisotropies of the cosmic microwave background (CMB) gave a strong tension with the local measurement from R11's result at 2.4$\sigma$ confidence level. In order to resolve or alleviate the tension, several groups carried out the measurements utilizing different techniques and methods: Bennett et al. \cite{8} and Hinshaw et al. \cite{9} gave a $3\%$ determination, i.e., $H_0=70.0\pm2.2$ km s$^{-1}$ Mpc$^{-1}$ by using the nine-year Wilkinson Microwave Anisotropy Probe (WMAP-9) measurements; Spergel et al. \cite{10} found $H_0=68.0\pm1.1$ km s$^{-1}$ Mpc$^{-1}$ by removing the $217\times217$ GHz detector set spectrum used in the Planck analysis; Fiorentino et al. \cite{11} obtained $H_0=76.0\pm1.9$ km s$^{-1}$ Mpc$^{-1}$ by utilizing 8 new classical Cepheids observed in galaxies hosting SNe Ia; Different from the calibration method exhibited by R11, Tammann et al. \cite{12} got a lower value $H_0=63.7\pm2.3$ km s$^{-1}$ Mpc$^{-1}$ by calibrating the SN Ia with the tip of red-giant branch (TRGB); Efstathiou \cite{13} obtained $H_0=70.6\pm3.3$ km s$^{-1}$ Mpc$^{-1}$ by revising the geometric maser distance to NGC 4258 from Humphreys et al. \cite{14} and using this indicator to calibrate the R11's measurements; Rigault et al. \cite{15} gave $H_0=70.6\pm2.6$ km s$^{-1}$ Mpc$^{-1}$ by considering predominately star-forming environments.

The medium redshift data can also act as an  complementary and effective tool to determine $H_0$. Combining them with the high-redshift CMB data, the uncertainties of $H_0$ can be reduced obviously. By making the best of baryon acoustic oscillations (BAO) data, Cheng et al. \cite{16} found $H_0=68.0\pm1.1$ km s$^{-1}$ Mpc$^{-1}$ for the standard cosmological model. Utilizing the CMB and BAO data and assuming the six-parameter standard cosmology, Bennett et al. \cite{17} got a substantially accurate result $H_0=69.6\pm0.7$ km s$^{-1}$ Mpc$^{-1}$. Subsequently, using other mid-redshift data including the BAO peak at $z=0.35$ \cite{18}, 18 $H(z)$ data points \cite{19,20,21}, 11 ages of old high-redshift galaxies \cite{22,23} and the angular diameter distance data from the Bonamente et al. galaxy cluster sample \cite{24}, Lima et al. obtained  $H_0=74.1\pm2.2$ km s$^{-1}$ Mpc$^{-1}$ \cite{25} in a $\Lambda$CDM model. Furthermore, replacing the Bonamente et al. galaxy cluster sample with the Filippis et al. one \cite{26}, Holanda et al. \cite{27} gave $H_0=70\pm4$ km s$^{-1}$ Mpc$^{-1}$. This indicates that different mid-redshift data can provide different values of $H_0$. Moreover, based on the fact that different observations should provide the same luminosity distance at some certain redshift, Wu et al. \cite{28} proposed a model-independent method to determine $H_0$. They found $H_0=74.1\pm2.2$ km s$^{-1}$ Mpc$^{-1}$ by combining the Union 2.1 SNe Ia data with galaxy cluster data \cite{29}.

Recently, the improved local measurement $H_0=73.24\pm1.74$ km s$^{-1}$ Mpc$^{-1}$ from Riess et al. 2016 \cite{30} gives a stronger tension with the Planck 2016 release $H_0=66.93\pm0.62$ km s$^{-1}$ Mpc$^{-1}$ \cite{31} (hereafter P16) at $3.4\sigma$ confidence level. The improvements different from R11's result are summarized in the following manner:

$\star$ adopting new, near-infrared observations of Cepheid variables in 11 SNe Ia hosts; 

$\star$ increasing the sample size of ideal SNe Ia calibrators from 8 to 19; 

$\star$  giving the calibration for a magnitude-redshift relation based on 300 SNe Ia at $z < 0.15$; 

$\star$ a $33\%$ reduction of the systematic uncertainty in the maser distance to the NGC 4258;

$\star$ increasing the sample size of Cepheids in the LMC; 

$\star$ HST observations of Cepheids in M31; 

$\star$ using new HST-based trigonometric parallaxes for the MW Cepheids;

$\star$ a more robust distance to the Large Magellanic Cloud (LMC) based on the late-type detached eclipsing binaries (DEBs). 

In light of the higher tension than before between the local and global measurements of $H_0$, we use the latest HII galaxy data to determine the Hubble constant here. We find that the $H_0$ values are more consistent with the R16's result at $1\sigma$ confidence level by utilizing a combination of model-dependent and model-independent methods, and that the data prefers a higher underlying value of $H_0$ than R16's result.   

The rest of this paper is outlined as follows. In the next section, we describe the data we use in this analysis. In Sec. 3, we constrain five cosmological models. In Sec. 4, we use the model-independent GP reconstructions to check the correctness of the $H_0$ values from the model-dependent method. The discussions and conclusions are presented in the final section (we use units $8\pi G=c=1$).

\section{The HII galaxy measurements}
Our sample consists of 24 Giant Extragalactic HII Regions (GEHR) at redshifts $z\leqslant0.01$ analyzed in \cite{H6}, and 107 low-$z$ HII galaxies vividly described in \cite{H5}. This compilation also includes 6 high-$z$ star-forming galaxies in the redshift range $z\in[0.64, 2.33]$ obtained via the X-Shooter spectrography at the Cassegrain focus of the European Southern Observatory Very Large Telescope (ESO-VLT) \cite{C}. We also add 19 high-$z$ objects---1 from \cite{H3}, 6 from \cite{H1} and 12 from \cite{H4}---into our sample, therefore, we have 25 high-$z$ HII galaxy measurements \cite{H2}. In total, our curent sample includes 156 objects which are shown in Table. \ref{t1}
\begin{center}
\begin{longtable}{cccc}
\caption{The measured gas velocity dispersion and flux of GEHR and HII galaxies.} \label{t1} \\
\hline
\hline \multicolumn{1}{c}{$z$} & \multicolumn{1}{c}{$log\sigma(H\beta)$} & \multicolumn{1}{c}{$logf(H\beta)$} & \multicolumn{1}{c}{Ref.} \\ \hline
\endfirsthead

\multicolumn{3}{c}%
{{ continued from previous page}} \\
\hline \multicolumn{1}{c}{$z$} & \multicolumn{1}{c}{$log\sigma(H\beta)$} & \multicolumn{1}{c}{$logf(H\beta)$} & \multicolumn{1}{c}{Ref.}  \\ \hline
\endhead

\hline \multicolumn{3}{r}{{Continued on next page}} \\ \hline
\endfoot

\hline \hline
\endlastfoot
\hline
&                            \qquad\qquad\qquad GEHR                                                        \\
\hline
$0.00012$        &$1.013\pm0.035$    &$-11.131\pm0.102$      & \cite{H6}               \\
$0.00012$        &$1.021\pm0.035$    &$-11.137\pm0.095$      & \cite{H6}               \\
$0.00001$        &$1.061\pm0.035$    &$-9.083\pm0.095$      & \cite{H6}               \\
$0.00020$        &$1.111\pm0.035$    &$-11.269\pm0.095$      & \cite{H6}               \\
$0.00110$        &$1.133\pm0.036$    &$-12.509\pm0.102$      & \cite{H6}               \\
$0.00110$        &$1.159\pm0.035$    &$-12.181\pm0.102$      & \cite{H6}               \\
$0.00085$        &$1.176\pm0.035$    &$-11.953\pm0.102$      & \cite{H6}               \\
$0.00100$        &$1.199\pm0.035$    &$-12.185\pm0.095$      & \cite{H6}               \\
$0.00077$        &$1.204\pm0.035$    &$-12.101\pm0.095$      & \cite{H6}               \\
$0.00020$        &$1.201\pm0.035$    &$-11.082\pm0.095$      & \cite{H6}               \\
$0.00020$        &$1.250\pm0.036$    &$-10.733\pm0.102$      & \cite{H6}               \\
$0.00100$        &$1.250\pm0.036$    &$-12.232\pm0.095$      & \cite{H6}               \\
$0.00185$        &$1.267\pm0.035$    &$-12.619\pm0.095$      & \cite{H6}               \\
$0.00085$        &$1.207\pm0.035$    &$-11.571\pm0.095$      & \cite{H6}               \\
$0.00077$        &$1.267\pm0.035$    &$-11.579\pm0.095$      & \cite{H6}               \\
$0.00020$        &$1.277\pm0.035$    &$-10.285\pm0.095$      & \cite{H6}               \\
$0.00185$        &$1.293\pm0.035$    &$-12.278\pm0.102$      & \cite{H6}               \\
$0.00077$        &$1.320\pm0.035$    &$-11.713\pm0.102$      & \cite{H6}               \\
$0.00001$        &$1.369\pm0.035$    &$-7.959\pm0.095$      & \cite{H6}               \\
$0.00077$        &$1.384\pm0.035$    &$-11.258\pm0.102$      & \cite{H6}               \\
$0.00185$        &$1.314\pm0.035$    &$-11.983\pm0.095$      & \cite{H6}               \\
$0.00185$        &$1.310\pm0.035$    &$-11.775\pm0.095$      & \cite{H6}               \\
$0.00185$        &$1.333\pm0.035$    &$-11.695\pm0.095$      & \cite{H6}               \\
$0.00185$        &$1.351\pm0.035$    &$-11.722\pm0.095$      & \cite{H6}               \\

\hline
&                            \qquad\qquad\qquad\qquad low-$z$ HII galaxies \\
\hline
$0.02203$        &$1.377\pm0.039$    &$-13.096\pm0.141$      & \cite{H5}               \\
$0.05191$        &$1.463\pm0.036$    &$-13.411\pm0.120$      & \cite{H5}               \\
$0.01257$        &$1.538\pm0.034$    &$-13.229\pm0.049$      & \cite{H5}               \\
$0.05191$        &$1.463\pm0.036$    &$-13.411\pm0.120$      & \cite{H5}               \\
$0.03637$        &$1.454\pm0.036$    &$-13.392\pm0.049$      & \cite{H5}               \\
$0.05712$        &$1.529\pm0.034$    &$-13.954\pm0.098$      & \cite{H5}               \\
$0.09424$        &$1.527\pm0.033$    &$-14.119\pm0.049$      & \cite{H5}               \\
$0.01812$        &$1.283\pm0.042$    &$-13.987\pm0.062$      & \cite{H5}               \\
$0.01718$        &$1.369\pm0.040$    &$-13.610\pm0.109$      & \cite{H5}               \\
$0.05574$        &$1.625\pm0.033$    &$-13.603\pm0.075$      & \cite{H5}               \\
$0.01207$        &$1.144\pm0.060$    &$-13.724\pm0.120$      & \cite{H5}               \\
$0.11235$        &$1.706\pm0.033$    &$-13.671\pm0.075$      & \cite{H5}               \\
$0.08164$        &$1.651\pm0.034$    &$-13.554\pm0.075$      & \cite{H5}               \\
$0.07687$        &$1.590\pm0.034$    &$-13.822\pm0.062$      & \cite{H5}               \\
$0.01420$        &$1.390\pm0.038$    &$-13.575\pm0.049$      & \cite{H5}               \\
$0.16417$        &$1.782\pm0.032$    &$-14.041\pm0.049$      & \cite{H5}               \\
$0.05097$        &$1.419\pm0.039$    &$-14.025\pm0.062$      & \cite{H5}               \\
$0.02282$        &$1.350\pm0.041$    &$-13.757\pm0.086$      & \cite{H5}               \\
$0.07443$        &$1.548\pm0.034$    &$-13.934\pm0.062$      & \cite{H5}               \\
$0.05041$        &$1.446\pm0.026$    &$-13.505\pm0.255$      & \cite{H5}               \\
$0.06347$        &$1.576\pm0.025$    &$-13.635\pm0.109$      & \cite{H5}               \\
$0.07485$        &$1.567\pm0.022$    &$-14.009\pm0.098$      & \cite{H5}               \\
$0.03993$        &$1.484\pm0.026$    &$-13.653\pm0.075$      & \cite{H5}               \\
$0.07051$        &$1.791\pm0.032$    &$-13.187\pm0.062$      & \cite{H5}               \\
$0.02021$        &$1.463\pm0.035$    &$-13.331\pm0.098$      & \cite{H5}               \\
$0.02077$        &$1.480\pm0.026$    &$-13.755\pm0.075$      & \cite{H5}               \\
$0.05547$        &$1.565\pm0.033$    &$-13.808\pm0.062$      & \cite{H5}               \\
$0.02370$        &$1.588\pm0.033$    &$-13.230\pm0.141$      & \cite{H5}               \\
$0.08769$        &$1.532\pm0.035$    &$-14.116\pm0.109$      & \cite{H5}               \\
$0.09729$        &$1.649\pm0.033$    &$-13.736\pm0.062$      & \cite{H5}               \\
$0.10937$        &$1.688\pm0.025$    &$-13.538\pm0.120$      & \cite{H5}               \\
$0.11245$        &$1.707\pm0.024$    &$-13.852\pm0.086$      & \cite{H5}               \\
$0.07302$        &$1.664\pm0.019$    &$-13.767\pm0.062$      & \cite{H5}               \\
$0.04258$        &$1.637\pm0.034$    &$-13.169\pm0.086$      & \cite{H5}               \\
$0.08479$        &$1.652\pm0.024$    &$-13.587\pm0.086$      & \cite{H5}               \\
$0.03065$        &$1.490\pm0.035$    &$-13.113\pm0.120$      & \cite{H5}               \\
$0.09209$        &$1.747\pm0.024$    &$-13.289\pm0.109$      & \cite{H5}               \\
$0.03127$        &$1.449\pm0.035$    &$-13.451\pm0.086$      & \cite{H5}               \\
$0.01125$        &$1.406\pm0.040$    &$-13.023\pm0.098$      & \cite{H5}               \\
$0.07687$        &$1.725\pm0.032$    &$-13.152\pm0.075$      & \cite{H5}               \\
$0.09922$        &$1.766\pm0.024$    &$-13.513\pm0.086$      & \cite{H5}               \\
$0.12641$        &$1.646\pm0.025$    &$-13.894\pm0.062$      & \cite{H5}               \\
$0.04038$        &$1.566\pm0.025$    &$-13.763\pm0.049$      & \cite{H5}               \\
$0.02771$        &$1.535\pm0.035$    &$-13.156\pm0.033$      & \cite{H5}               \\
$0.02293$        &$1.477\pm0.035$    &$-13.651\pm0.075$      & \cite{H5}               \\
$0.05815$        &$1.614\pm0.024$    &$-13.919\pm0.086$      & \cite{H5}               \\
$0.10809$        &$1.737\pm0.023$    &$-13.668\pm0.120$      & \cite{H5}               \\
$0.09494$        &$1.561\pm0.025$    &$-13.915\pm0.086$      & \cite{H5}               \\
$0.01352$        &$1.441\pm0.036$    &$-13.232\pm0.049$      & \cite{H5}               \\
$0.08536$        &$1.700\pm0.024$    &$-13.693\pm0.075$      & \cite{H5}               \\
$0.04038$        &$1.566\pm0.025$    &$-13.763\pm0.049$      & \cite{H5}               \\
$0.10263$        &$1.787\pm0.031$    &$-13.343\pm0.062$      & \cite{H5}               \\
$0.04587$        &$1.602\pm0.025$    &$-14.003\pm0.033$      & \cite{H5}               \\
$0.01558$        &$1.536\pm0.034$    &$-13.485\pm0.049$      & \cite{H5}               \\
$0.02329$        &$1.496\pm0.036$    &$-14.095\pm0.086$      & \cite{H5}               \\
$0.01765$        &$1.434\pm0.036$    &$-13.534\pm0.049$      & \cite{H5}               \\
$0.01822$        &$1.440\pm0.035$    &$-13.541\pm0.075$      & \cite{H5}               \\
$0.09883$        &$1.750\pm0.025$    &$-13.640\pm0.086$      & \cite{H5}               \\
$0.12029$        &$1.746\pm0.025$    &$-13.591\pm0.086$      & \cite{H5}               \\
$0.01578$        &$1.296\pm0.044$    &$-13.516\pm0.062$      & \cite{H5}               \\
$0.01621$        &$1.425\pm0.042$    &$-13.220\pm0.109$      & \cite{H5}               \\
$0.03259$        &$1.297\pm0.035$    &$-14.176\pm0.086$      & \cite{H5}               \\
$0.02518$        &$1.532\pm0.034$    &$-13.372\pm0.086$      & \cite{H5}               \\
$0.06244$        &$1.681\pm0.042$    &$-13.261\pm0.049$      & \cite{H5}               \\
$0.05564$        &$1.612\pm0.025$    &$-13.661\pm0.086$      & \cite{H5}               \\
$0.14486$        &$1.709\pm0.032$    &$-13.962\pm0.062$      & \cite{H5}               \\
$0.14807$        &$1.774\pm0.024$    &$-13.811\pm0.062$      & \cite{H5}               \\
$0.01390$        &$1.279\pm0.044$    &$-13.986\pm0.075$      & \cite{H5}               \\
$0.03476$        &$1.560\pm0.037$    &$-13.172\pm0.049$      & \cite{H5}               \\
$0.03375$        &$1.540\pm0.034$    &$-13.347\pm0.296$      & \cite{H5}               \\
$0.04583$        &$1.791\pm0.033$    &$-12.873\pm0.109$      & \cite{H5}               \\
$0.06988$        &$1.597\pm0.022$    &$-13.999\pm0.098$      & \cite{H5}               \\
$0.05050$        &$1.630\pm0.034$    &$-14.027\pm0.062$      & \cite{H5}               \\
$0.07806$        &$1.593\pm0.025$    &$-13.865\pm0.086$      & \cite{H5}               \\
$0.01453$        &$1.410\pm0.038$    &$-13.365\pm0.049$      & \cite{H5}               \\
$0.02777$        &$1.593\pm0.034$    &$-13.037\pm0.075$      & \cite{H5}               \\
$0.01216$        &$1.446\pm0.036$    &$-13.260\pm0.062$      & \cite{H5}               \\
$0.03039$        &$1.639\pm0.033$    &$-12.656\pm0.109$      & \cite{H5}               \\
$0.08564$        &$1.561\pm0.033$    &$-13.326\pm0.062$      & \cite{H5}               \\
$0.05314$        &$1.544\pm0.026$    &$-13.561\pm0.062$      & \cite{H5}               \\
$0.04670$        &$1.569\pm0.034$    &$-13.909\pm0.075$      & \cite{H5}               \\
$0.15134$        &$1.587\pm0.032$    &$-14.180\pm0.075$      & \cite{H5}               \\
$0.12499$        &$1.660\pm0.024$    &$-13.974\pm0.062$      & \cite{H5}               \\
$0.01111$        &$1.396\pm0.038$    &$-13.372\pm0.086$      & \cite{H5}               \\
$0.02492$        &$1.434\pm0.027$    &$-13.463\pm0.062$      & \cite{H5}               \\
$0.02187$        &$1.465\pm0.016$    &$-13.360\pm0.086$      & \cite{H5}               \\
$0.02765$        &$1.407\pm0.020$    &$-13.364\pm0.275$      & \cite{H5}               \\
$0.02671$        &$1.431\pm0.022$    &$-13.434\pm0.120$      & \cite{H5}               \\
$0.02362$        &$1.309\pm0.032$    &$-13.567\pm0.062$      & \cite{H5}               \\
$0.01508$        &$1.424\pm0.027$    &$-13.407\pm0.049$      & \cite{H5}               \\
$0.03138$        &$1.609\pm0.024$    &$-13.234\pm0.062$      & \cite{H5}               \\
$0.03328$        &$1.683\pm0.024$    &$-13.475\pm0.120$      & \cite{H5}               \\
$0.02808$        &$1.688\pm0.024$    &$-12.907\pm0.062$      & \cite{H5}               \\
$0.03437$        &$1.739\pm0.023$    &$-13.263\pm0.005$      & \cite{H5}               \\
$0.01094$        &$1.340\pm0.021$    &$-13.608\pm0.062$      & \cite{H5}               \\
$0.00880$        &$1.494\pm0.025$    &$-12.579\pm0.235$      & \cite{H5}               \\
$0.02711$        &$1.397\pm0.017$    &$-13.537\pm0.075$      & \cite{H5}               \\
$0.02662$        &$1.441\pm0.036$    &$-13.861\pm0.098$      & \cite{H5}               \\
$0.10880$        &$1.559\pm0.045$    &$-14.114\pm0.075$      & \cite{H5}               \\
$0.11506$        &$1.757\pm0.033$    &$-13.772\pm0.120$      & \cite{H5}               \\
$0.10726$        &$1.707\pm0.025$    &$-13.911\pm0.141$      & \cite{H5}               \\
$0.06551$        &$1.627\pm0.031$    &$-13.653\pm0.062$      & \cite{H5}               \\
$0.07928$        &$1.662\pm0.033$    &$-13.499\pm0.086$      & \cite{H5}               \\
$0.06094$        &$1.660\pm0.033$    &$-13.120\pm0.086$      & \cite{H5}               \\
$0.02283$        &$1.318\pm0.046$    &$-13.527\pm0.098$      & \cite{H5}               \\
$0.02873$        &$1.568\pm0.033$    &$-13.147\pm0.098$      & \cite{H5}               \\
$0.03278$        &$1.393\pm0.041$    &$-14.091\pm0.086$      & \cite{H5}               \\
$0.06479$        &$1.573\pm0.033$    &$-13.723\pm0.098$      & \cite{H5}               \\
\hline
&                          \qquad\qquad\qquad\qquad high-$z$ HII galaxies \\
\hline
$1.47740$        &$1.756\pm0.017$    &$-15.884\pm0.043$      & \cite{H1}               \\
$2.30520$        &$1.758\pm0.016$    &$-16.518\pm0.017$      & \cite{H1}               \\
$2.17350$        &$1.808\pm0.016$    &$-16.473\pm0.019$      & \cite{H2}               \\
$0.63640$        &$1.597\pm0.023$    &$-15.791\pm0.177$      & \cite{H2}               \\
$0.85100$        &$1.695\pm0.049$    &$-15.801\pm0.177$      & \cite{H2}               \\
$0.68160$        &$1.527\pm0.027$    &$-15.960\pm0.175$      & \cite{H1}               \\
$2.27130$        &$1.799\pm0.062$    &$-16.637\pm0.038$      & \cite{H1}               \\
$2.16630$        &$1.792\pm0.070$    &$-17.073\pm0.018$      & \cite{H1}               \\
$2.18140$        &$1.845\pm0.093$    &$-16.641\pm0.055$      & \cite{H1}               \\
$2.18160$        &$1.785\pm0.028$    &$-16.042\pm0.099$      & \cite{H1}               \\
$2.26640$        &$1.792\pm0.084$    &$-16.727\pm0.025$      & \cite{H1}               \\
$2.03000$        &$1.699\pm0.035$    &$-16.300\pm0.007$      & \cite{H3}               \\
$1.41200$        &$1.664\pm0.084$    &$-16.832\pm0.427$      & \cite{H4}               \\
$1.30000$        &$1.686\pm0.045$    &$-16.264\pm0.042$      & \cite{H4}               \\
$1.44400$        &$1.834\pm0.045$    &$-16.377\pm0.055$      & \cite{H4}               \\
$1.50400$        &$1.839\pm0.066$    &$-16.273\pm0.146$      & \cite{H4}               \\
$1.54300$        &$1.641\pm0.040$    &$-16.475\pm0.045$      & \cite{H4}               \\
$1.61000$        &$1.746\pm0.039$    &$-16.461\pm0.038$      & \cite{H4}               \\
$2.15800$        &$1.814\pm0.040$    &$-16.372\pm0.052$      & \cite{H4}               \\
$2.17700$        &$1.830\pm0.039$    &$-16.456\pm0.041$      & \cite{H4}               \\
$2.19100$        &$1.765\pm0.063$    &$-16.899\pm0.043$      & \cite{H4}               \\
$2.21500$        &$1.628\pm0.041$    &$-16.927\pm0.017$      & \cite{H4}               \\
$2.23400$        &$1.702\pm0.043$    &$-16.562\pm0.041$      & \cite{H4}               \\
$2.26400$        &$1.838\pm0.044$    &$-16.233\pm0.037$      & \cite{H4}               \\
$2.31500$        &$1.693\pm0.044$    &$-16.411\pm0.171$      & \cite{H4}               \\
\hline
\end{longtable}
\end{center}

\begin{table}[h!]
\caption{1$\sigma$ confidence range of free parameters of five different cosmological models.}
\begin{tabular}{ccccccc}

\hline
\hline
Parameter            &$\Lambda$CDM            &non-flat $\Lambda$CDM        &$\omega$CDM            &DV           &HDE                     \\
\hline
$H_0$                &$76.12^{+3.47}_{-3.44}$ &$75.74^{+3.45}_{-3.39}$      &$75.67^{+3.72}_{-3.40}$ &$76.10^{+3.36}_{-3.42}$  &$75.65^{+3.86}_{-3.55}$                          \\
$\Omega_{m0}$        &$0.265^{+0.176}_{-0.119}$ &$0.263^{+0.177}_{-0.121}$  &$0.51^{+0.189}_{-0.174}$ &$0.266^{+0.176}_{-0.120}$ &$0.271^{+0.134}_{-0.117}$                           \\
$\omega$             &---                        &---                      &$-0.844^{+0.360}_{-0.953}$           &---             &---                          \\
$\epsilon$           &---                        &---                      &---            &$0.010^{+0.005}_{-0.007}$             &---                          \\
$c$                  &---                        &---                      &---            &---             &$1.466^{+1.318}_{-0.920}$                         \\
\hline
$\chi^2_{min}$       &222.228                           &222.217                         &222.003               &222.222                &221.757    \\
\hline
\hline
\end{tabular}
\label{t2}
\end{table}

In [], the authors has pointed out that, for GEHR and HII galaxies, the $L(H\beta)-\sigma$ relation can be applied into measuring the distance, and it can be expressed as
\begin{equation}
logL(H\beta)=(5.05\pm0.097)log\sigma(H\beta)+(33.11\pm0.145), \label{1}
\end{equation}
where $L(H\beta)$ and $\sigma(H\beta)$ denote the Balmer emission line luminosity for these objects and the velocity dispersion of the young star-forming cluster from the measurements of the line width, respectively. Then the corresponding observational distance modulus is shown as
\begin{equation}
\mu_{obs}=2.5logL(H\beta)-2.5logf(H\beta)-100.95, \label{2}
\end{equation}
where $f(H\beta)$ is the measured flux in the $H\beta$ line. The theoretical distance modulus $\mu_{th}$ for a GEHR or HII galaxy can be written as
\begin{equation}
\mu_{th}=5\log_{10}d_L(z)+25. \label{3}
\end{equation}
The Hubble luminosity distance in a Friedmann-Robertson-Walker (FRW) universe is 
\begin{equation}
d_L(z)=\frac{1+z}{H_0\sqrt{|\Omega_{k0}|}}sinn\left(\sqrt{|\Omega_{k0}|}\int^{z}_{0}\frac{dz'}{E(z';\theta)}\right), \label{4}
\end{equation}
where $\theta$ denotes the model parameters, $H_0$ is the Hubble constant, the dimensionless Hubble parameter $E(z; \theta)=H(z; \theta)/H_0$, the present-day cosmic curvature $\Omega_{k0}=-K/(a_0H_0^2)$,  and for $sinn(x)= sin(x), x, sinh(x)$, $K=1, 0, -1$ , which corresponds to a closed, flat and open universe, respectively.

In order to constrain a specific cosmological model, we adopt the maximal likelihood method and the corresponding $\chi^2$ function to be minimized for the GEHR and HII galaxy data is
\begin{equation}
\chi^2=\sum^{156}_{i=1}[\frac{\mu_{obs}(z_i)-\mu_{th}(z_i;\theta)}{\sigma_i}]^2, \label{5}
\end{equation}
where $\sigma_i$ and $\mu_{obs}(z_i)$ are the $1\sigma$ statistical error and the observed value of distance modulus at a given redshift $z_i$ for every object, respectively.
\begin{figure}
\centering
\includegraphics[scale=0.27]{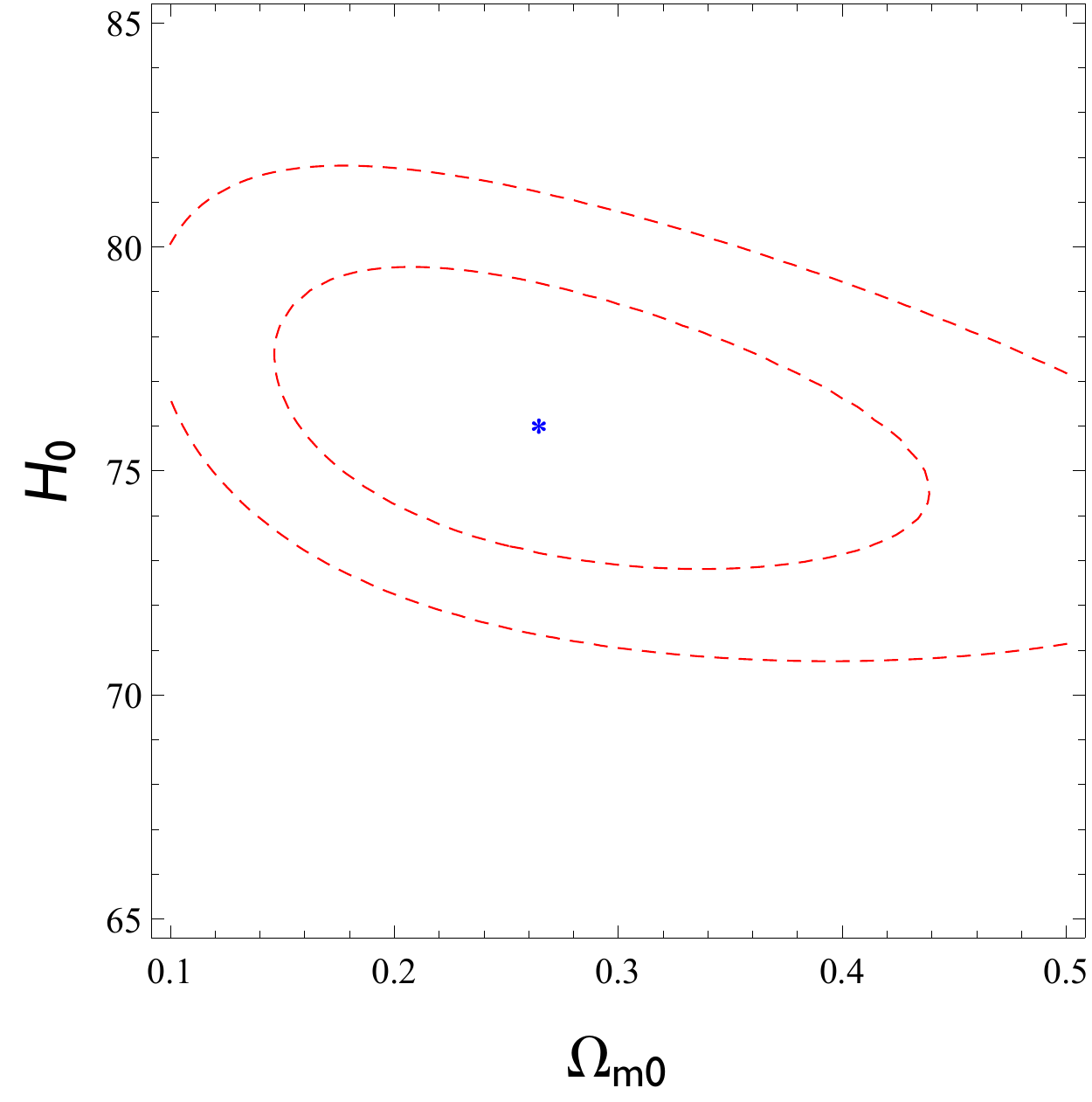}
\includegraphics[scale=0.27]{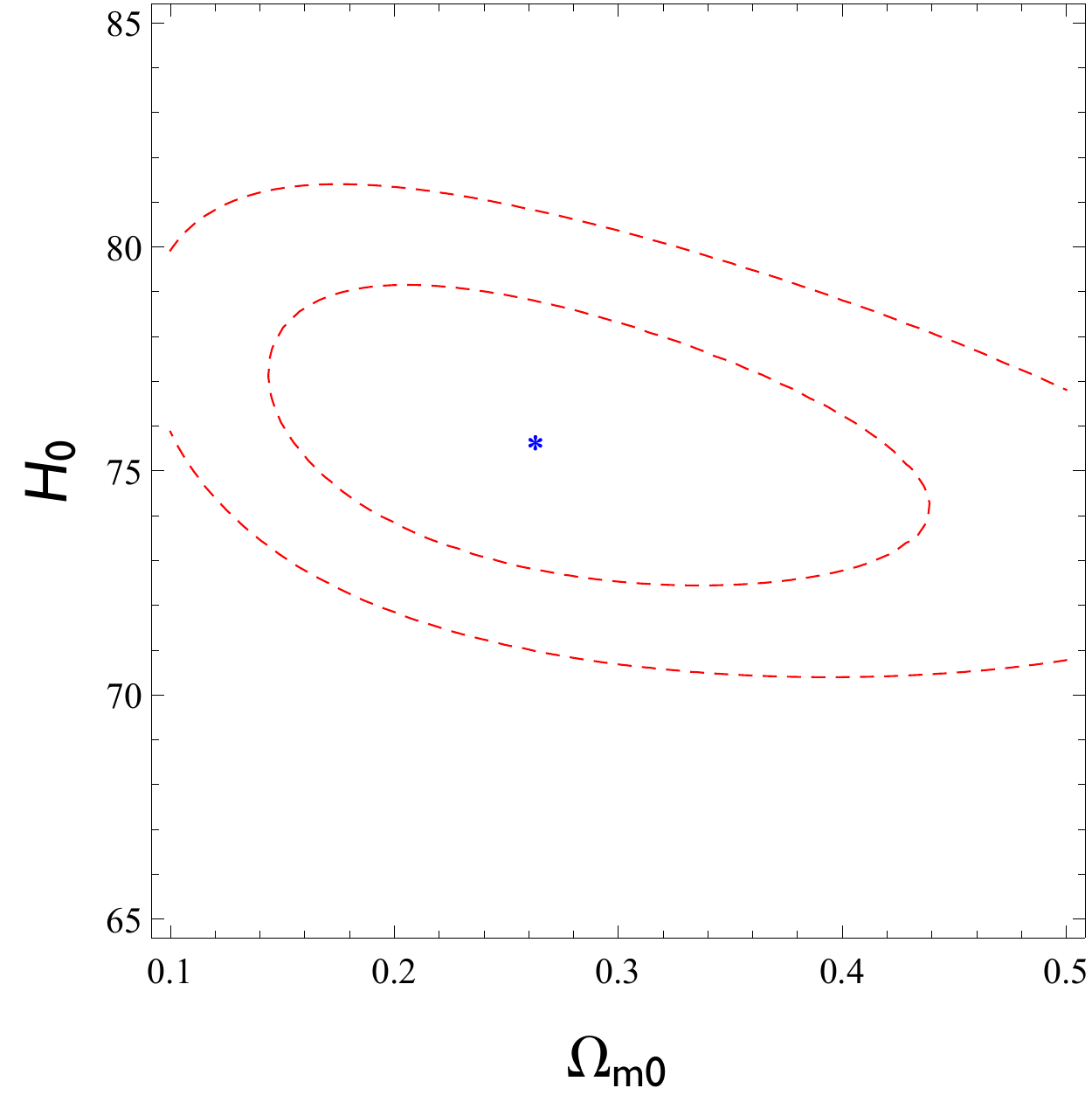}
\includegraphics[scale=0.27]{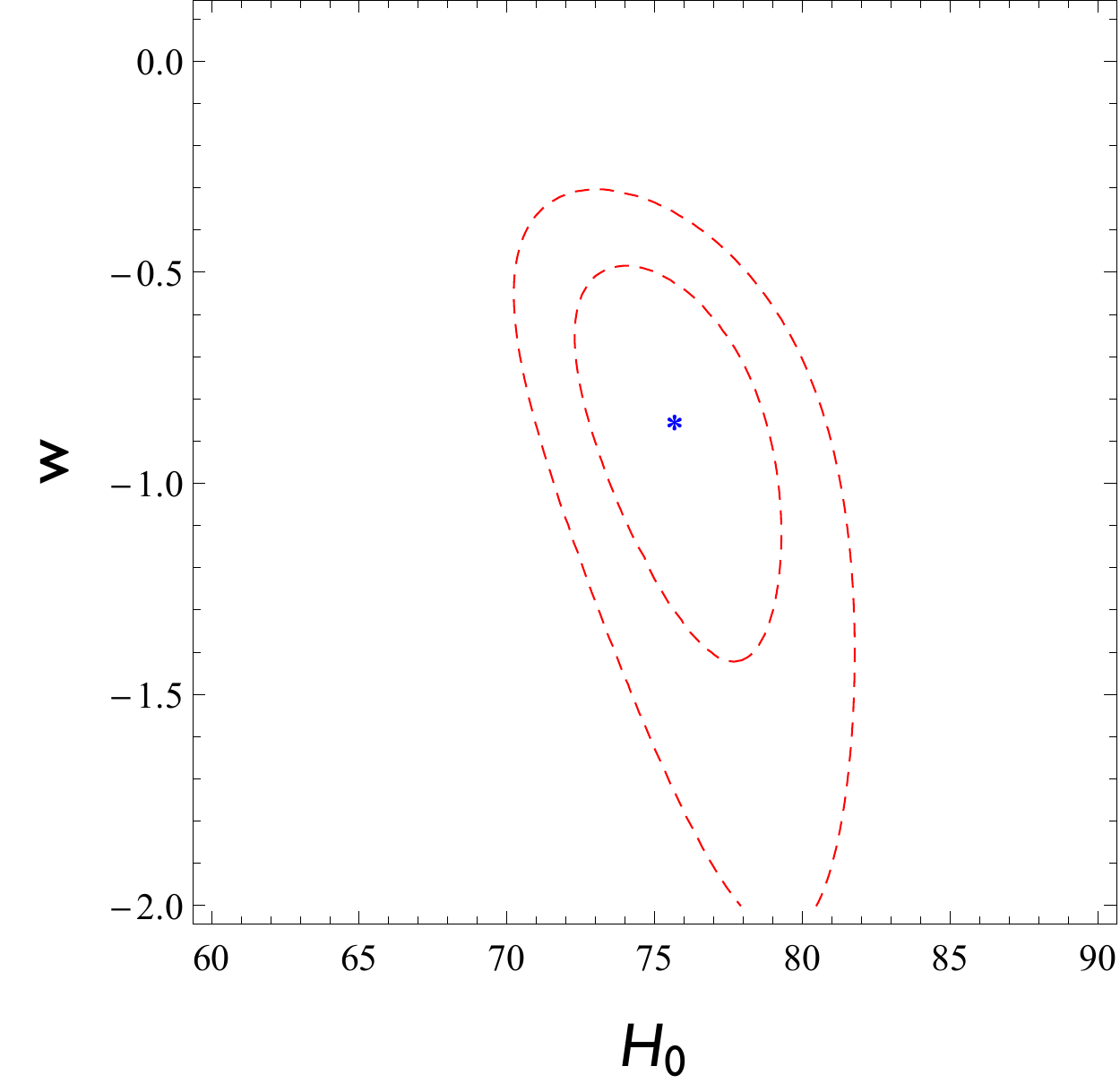}
\includegraphics[scale=0.27]{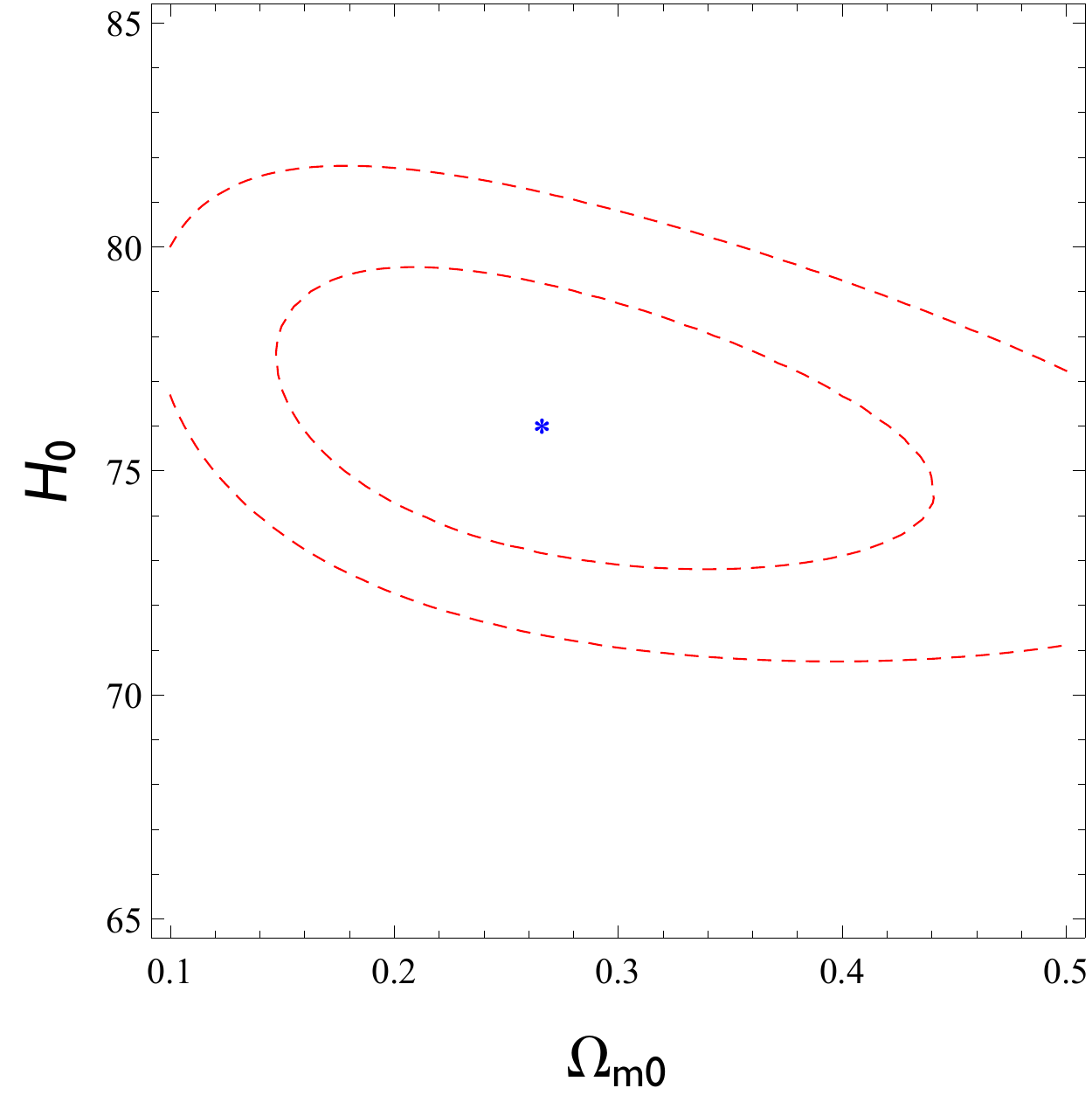}
\includegraphics[scale=0.27]{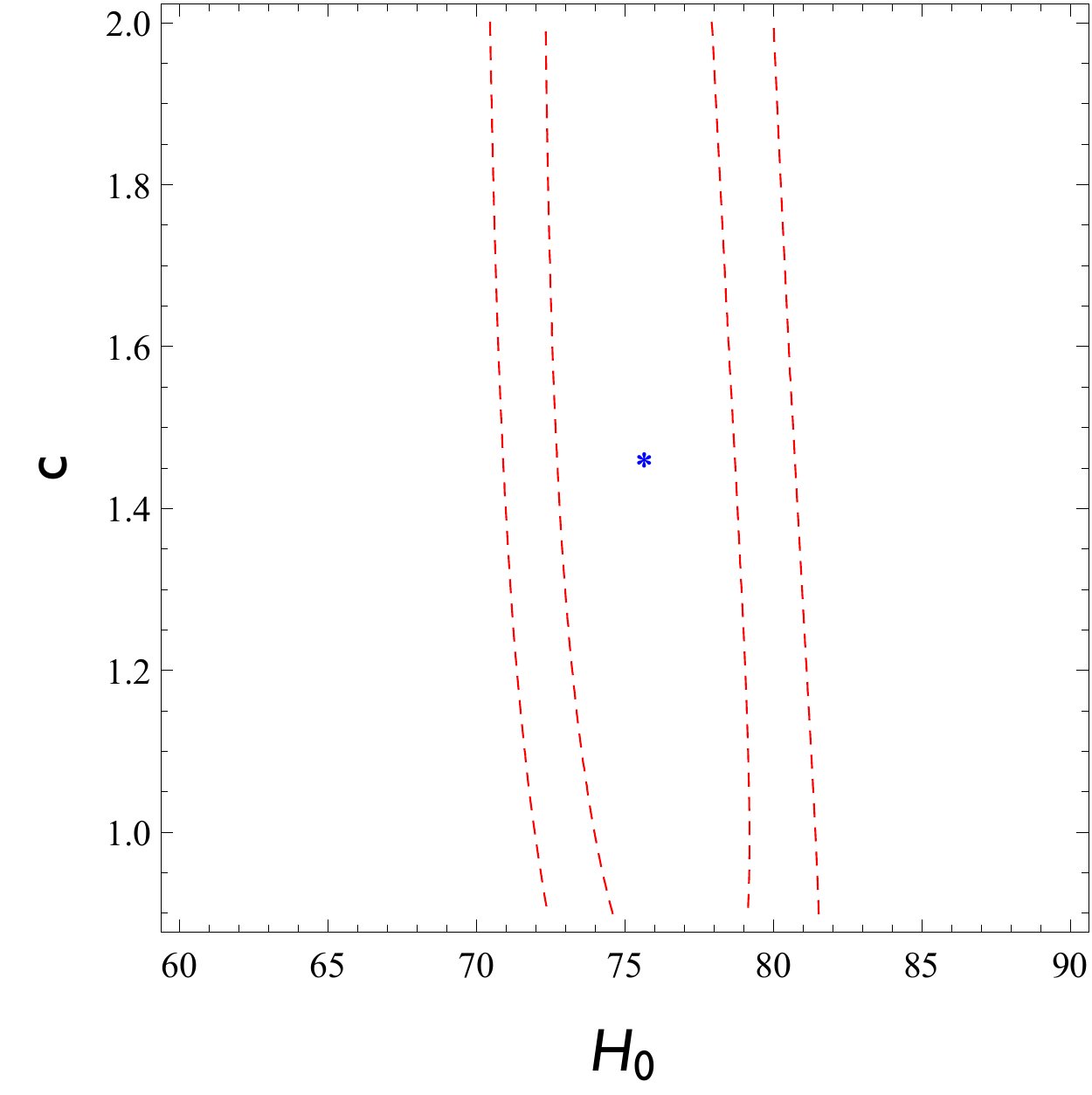}
\caption{1$\sigma$ and 2$\sigma$ confidence ranges for parameter pairs ($H_0$, $\Omega_{m0}$) of the $\Lambda$CDM model, ($H_0$, $\Omega_{m0}$) of the non-flat $\Lambda$CDM model, ($H_0$, $\omega$) of the $\omega$CDM model, ($H_0$, $\omega$) of the $\omega$CDM model, ($H_0$, $\Omega_{m0}$) of the DV model and ($H_0$, $c$) of the HDE model, respectively. The blue symbols `` $\ast$ '' represent the best-fit points.}\label{f1}
\end{figure}
\begin{figure}
\centering
\includegraphics[scale=0.4]{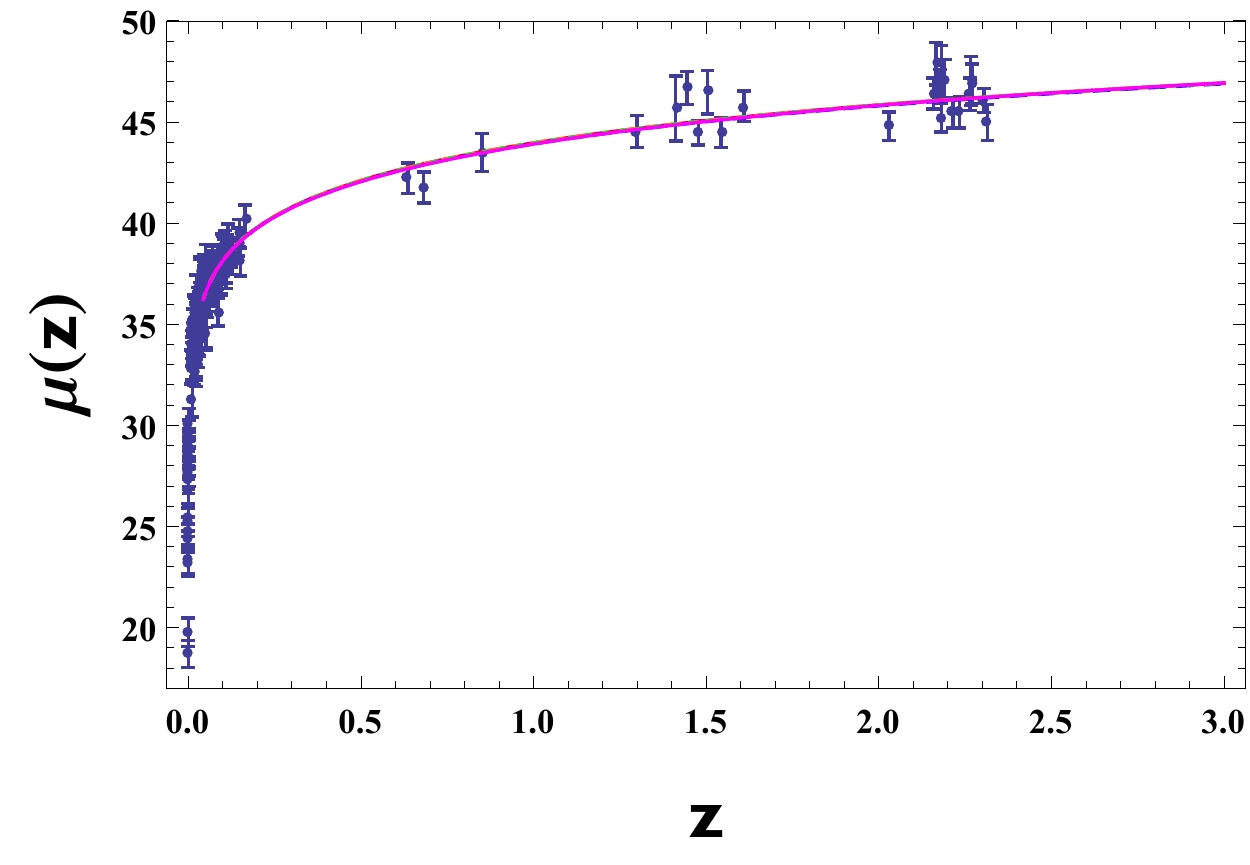}
\caption{The HII galaxies Hubble diagram for five cosmological models. The red (solid) line, blue (long-dashed) line, green (short-dashed) line, orange (dotted) line and magenta (dot-dashed) line correspond to the $\Lambda$CDM model, non-flat $\Lambda$CDM model, $\omega$CDM model, DV model and HDE model, respectively. }\label{f2}
\end{figure}
\begin{figure}
\centering
\includegraphics[scale=0.28]{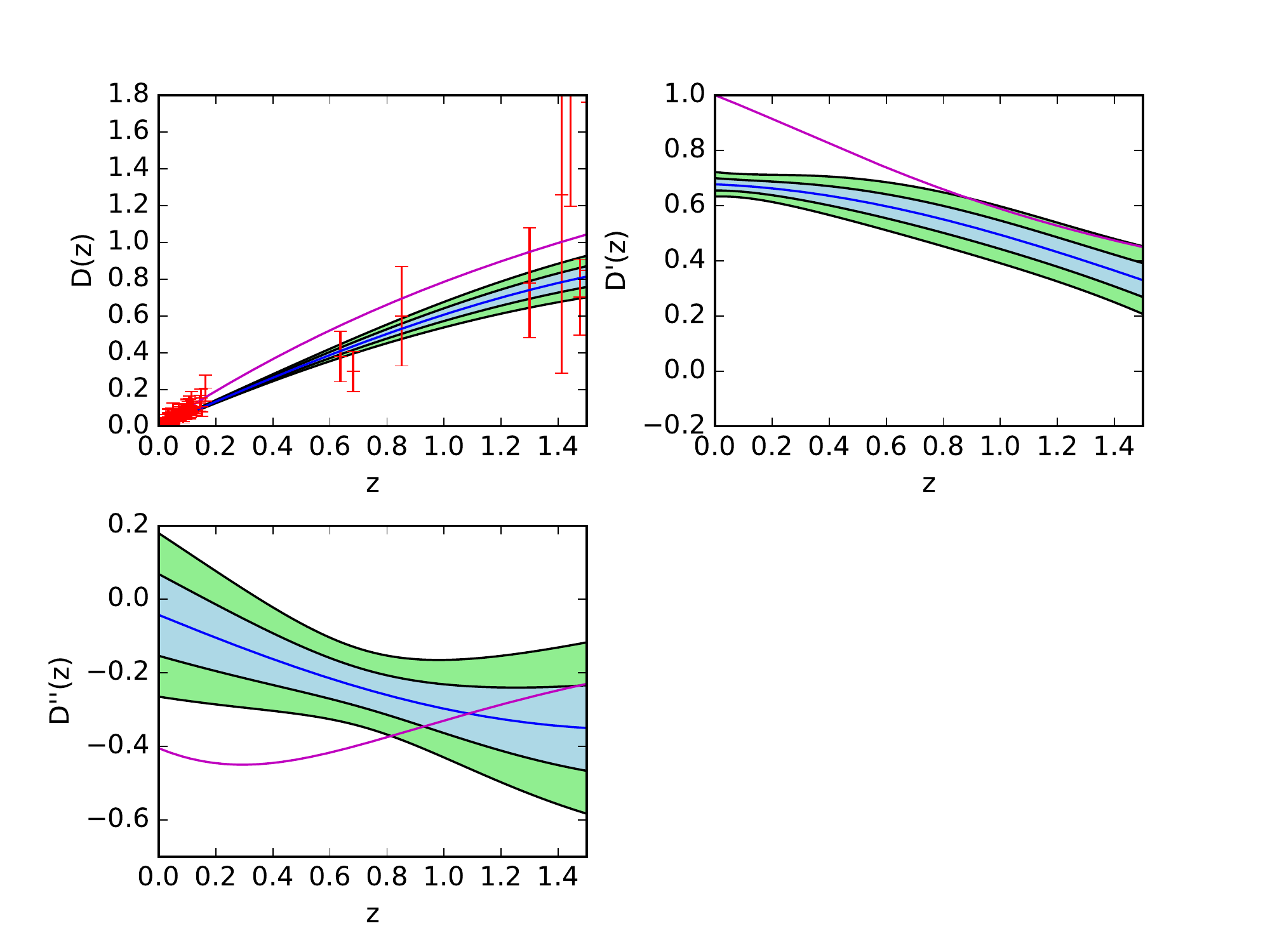}
\includegraphics[scale=0.28]{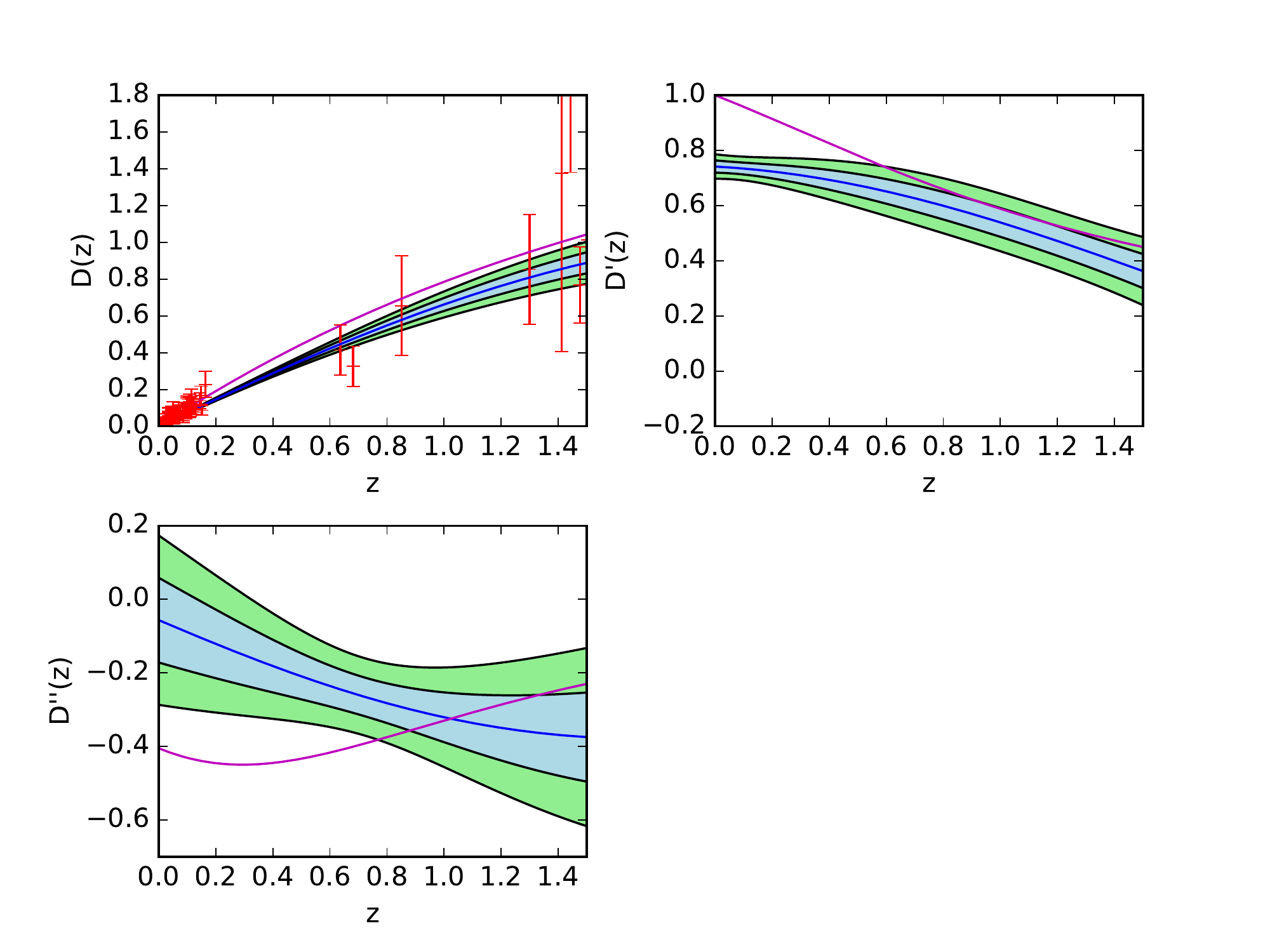}
\includegraphics[scale=0.28]{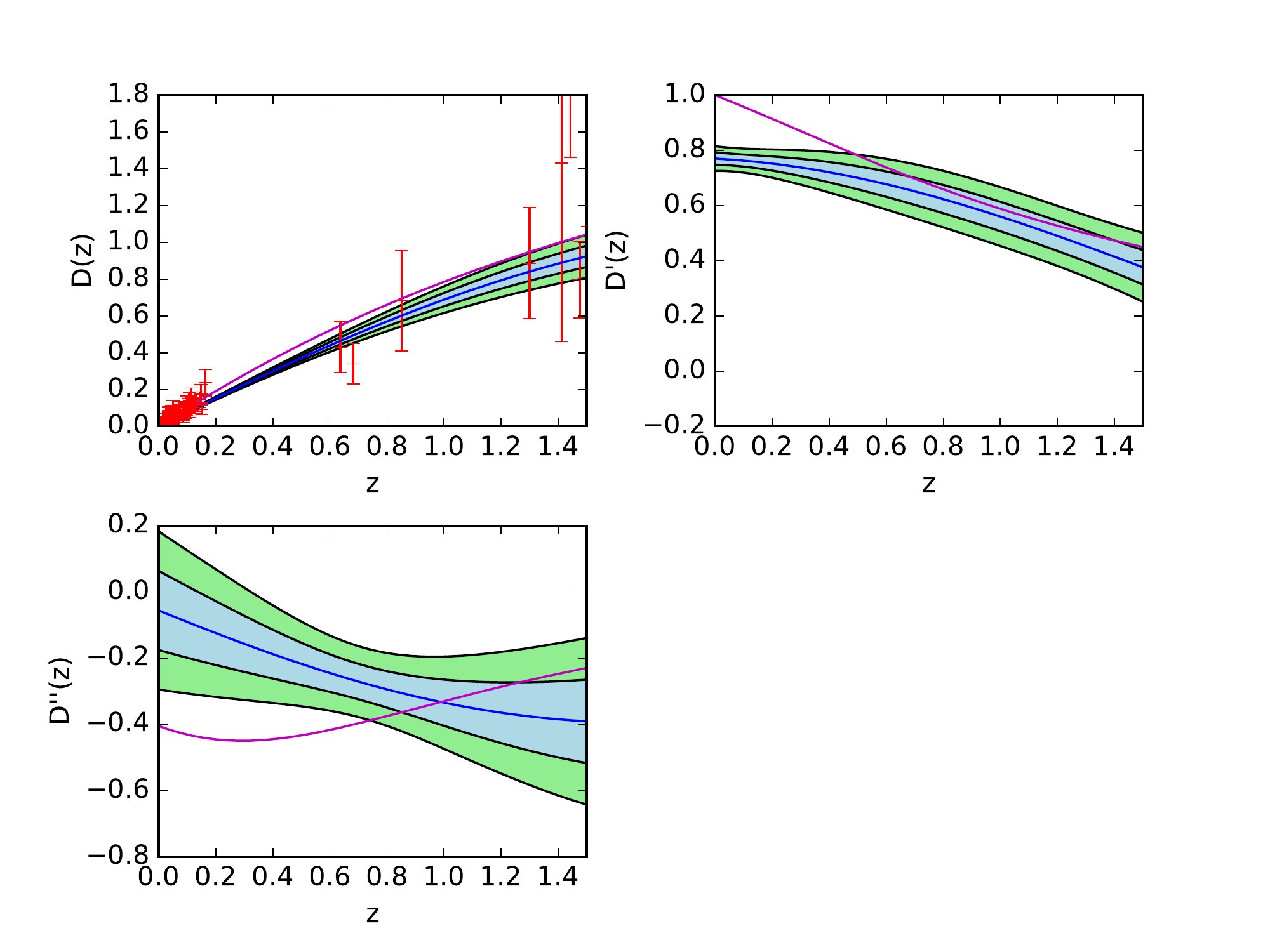}
\includegraphics[scale=0.28]{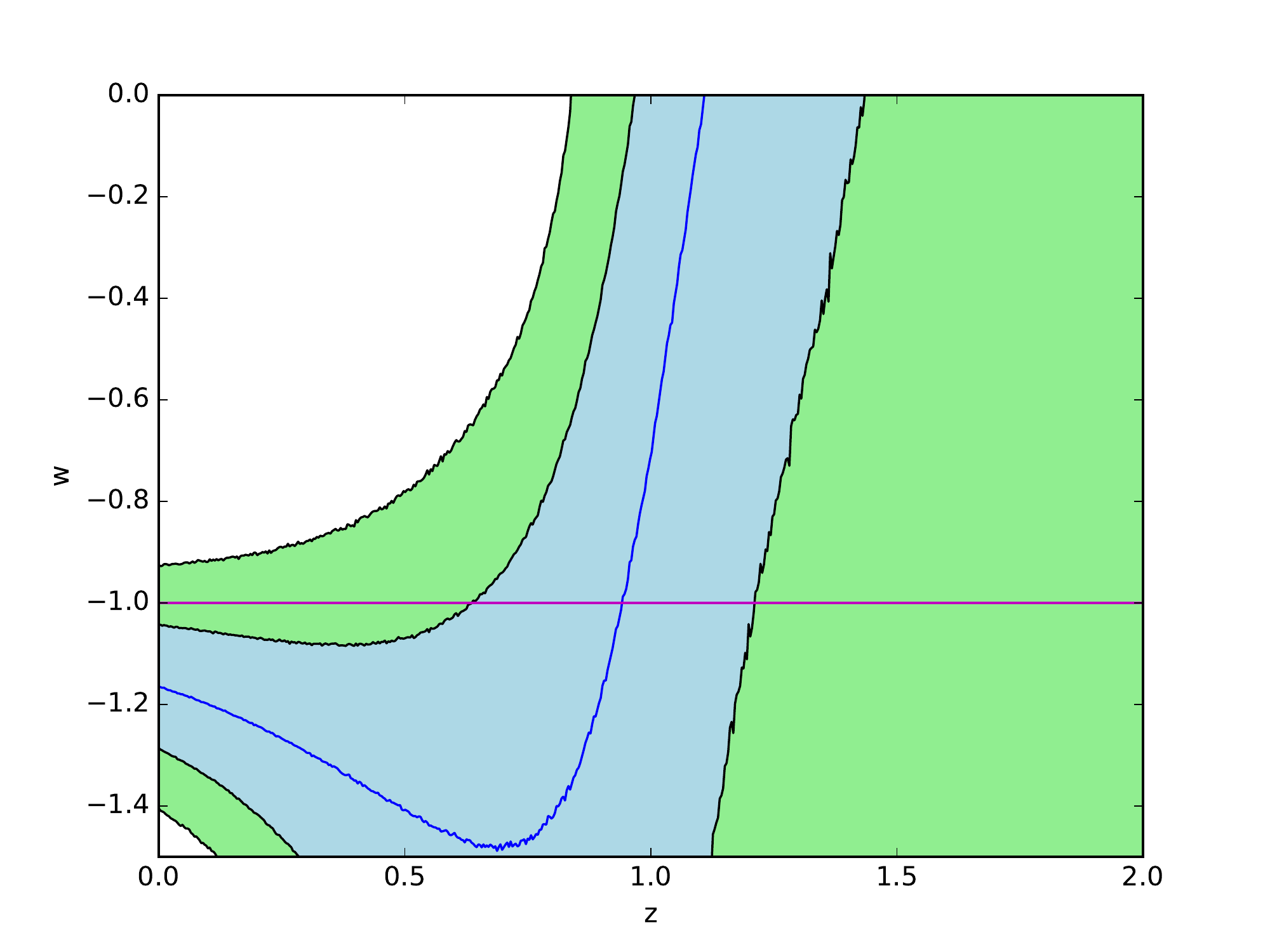}
\includegraphics[scale=0.28]{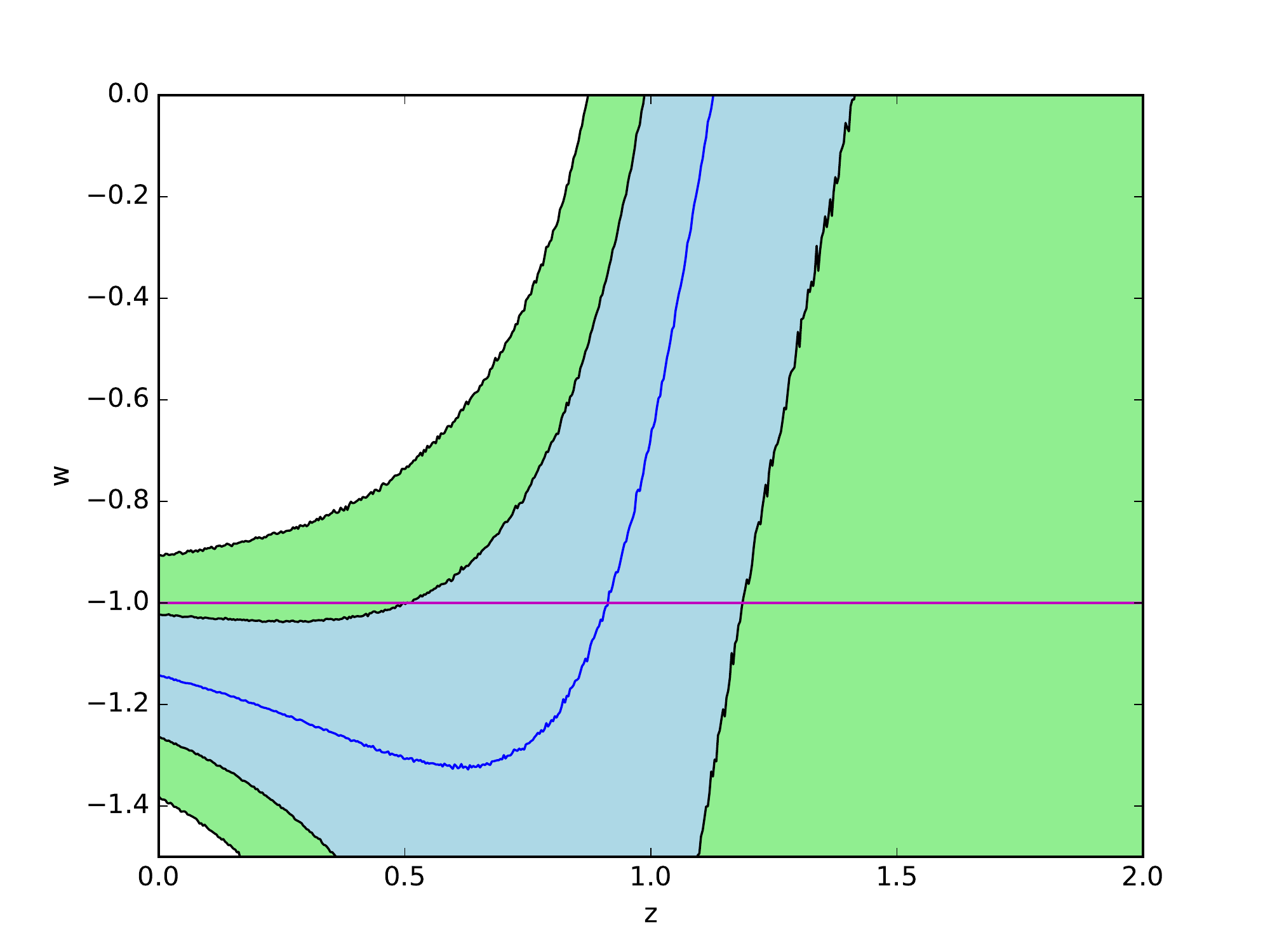}
\includegraphics[scale=0.28]{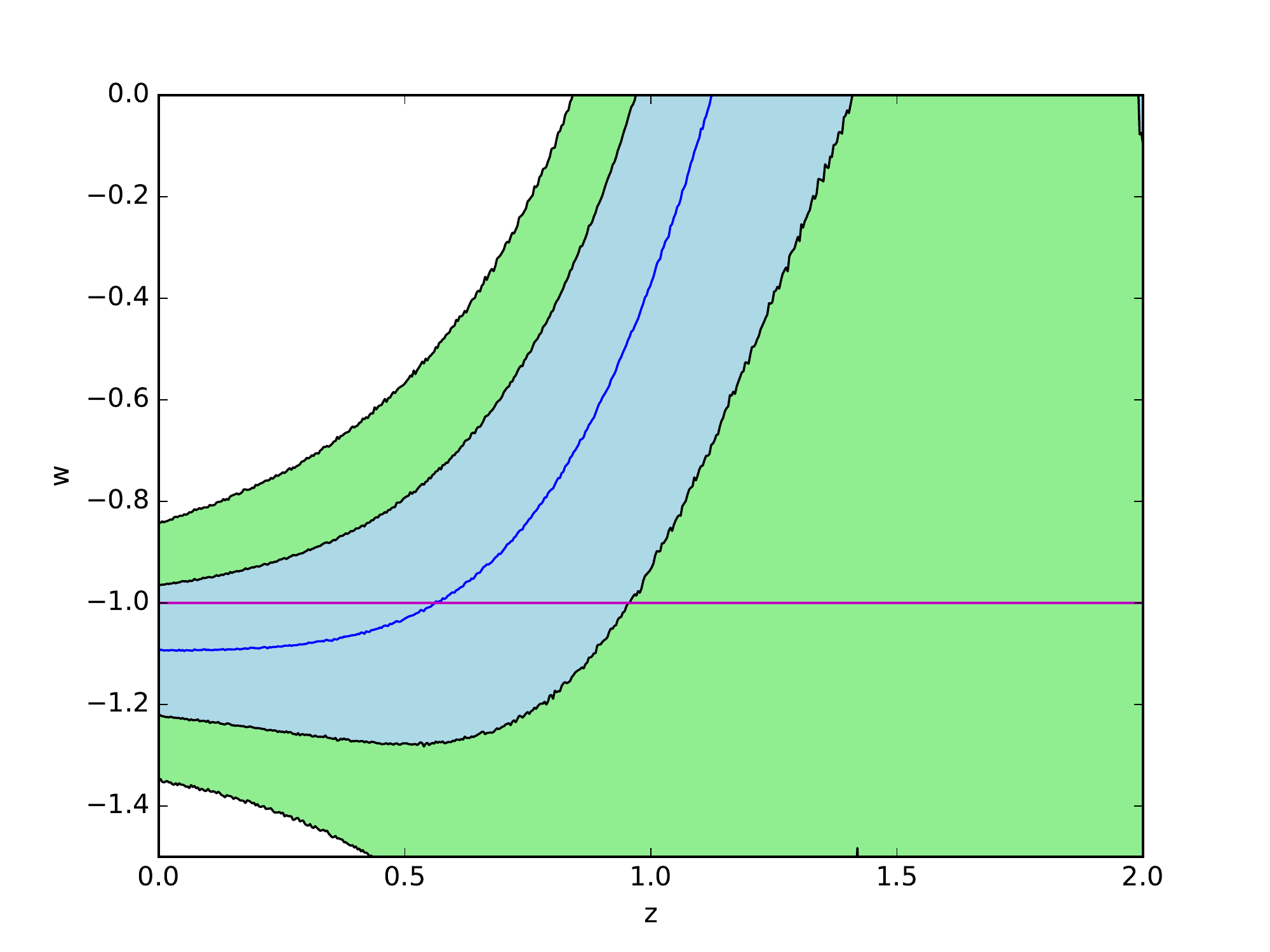}
\caption{Upper panels (from left to right): the reconstructions of $D(z), D'(z)$ and $D''(z)$ correspond to the cases of $H_0=66.93\pm0.62$, $73.24\pm1.74$ and $76.12^{+3.47}_{-3.44}$ km s$^{-1}$ Mpc$^{-1}$, respectively. Lower panels (from left to right): the reconstructions of the dark energy EoS correspond to the cases of $H_0=66.93\pm0.62$, $73.24\pm1.74$ and $76.12^{+3.47}_{-3.44}$ km s$^{-1}$ Mpc$^{-1}$, respectively. The points with red error bars are 156 HII galaxy measurements. The shaded regions are reconstructions with $68\%$ and $95\%$ confidence level. The blue lines and magenta lines represent the underlying true model (the mean value of reconstructions) and the $\Lambda$CDM model, respectively. }\label{f3}
\end{figure}

\section{The Constraints}
In this section, for the purpose to explore the values of $H_0$, we place constraints on five cosmological models by using the latest HII galaxy measurements. 

The Hubble parameter for the spatially flat $\Lambda$-cold-dark-matter ($\Lambda$CDM) model is 
\begin{equation}
H(z)=H_0\sqrt{\Omega_{m0}(1+z)^3+1-\Omega_{m0}}, \label{6}
\end{equation}  
while for the non-flat $\Lambda$CDM model it can be expressed as
\begin{equation}
H(z)=H_0\sqrt{\Omega_{m0}(1+z)^3+\Omega_{k0}(1+z)^2+1-\Omega_{m0}-\Omega_{k0}}, \label{7}
\end{equation}   
where $\Omega_{m0}$ and $\Omega_{k0}$ denote dimensionless matter density ratio parameter and the dimensionless curvature density ratio parameter today, respectively. 

We consider the simplest parameterization of dark energy equation of state (EoS) $\omega(z)=\omega=constant$, namely the $\omega$CDM model, and the corresponding Hubble parameter for the spatially flat $\omega$CDM model can be written as
\begin{equation}
H(z)=H_0\sqrt{\Omega_{m0}(1+z)^3+(1-\Omega_{m0})(1+z)^{3(1+\omega)}}. \label{8}
\end{equation}  

Another interesting model is the so-called decaying vacuum (DV) model, which is based on the a simple assumption about the form of the modified matter expansion rate \cite{39}. The Hubble parameter for the DV model is 
\begin{equation}
H(z)=H_0\sqrt{\frac{3\Omega_{m0}}{3-\epsilon}(1+z)^{3-\epsilon}+1-\frac{3\Omega_{m0}}{3-\epsilon}}. \label{9}
\end{equation}  
where $\epsilon$ is a small positive constant describing the deviation from the standard matter expansion rate. 

We also take into account the holographic dark energy (HDE) model inspired by the extraordinary thermodynamics of black holes \cite{40}, and the corresponding Hubble parameter can be shown as
\begin{equation}
H(z)=H_0\sqrt{\Omega_{m0}(1+z)^3+(1-\Omega_{m0})(1+z)^{2(1+\frac{1}{c}\sqrt{1-\Omega_{m0}})}}, \label{10}
\end{equation}  
where the parameter $c$ plays a main role in the HDE model. 

The minimal values of the derived $\chi^2$ and the best-fit values of the parameters for these five cosmological models are listed in Table. \ref{t2}. We also exhibit the $1\sigma$ and $2\sigma$ contour plots in Fig. \ref{f1}. One can easily find that, by using the latest HII galaxy data, the $H_0$ values for these five cosmological models are all consistent with the R16's local measurement at $1\sigma$ confidence level. In addition, we also conclude that these five models prefer a higher best-fit $H_0$ value than R16's result. Furthermore, we find that these five models fit the current data well and can not be distinguished from each other in the HII galaxies Hubble diagram (see Fig. \ref{f2}). This means that there exists a very high degeneracy among these five models using only the latest HII galaxy data.
 
\section{The GP reconstructions}
We have adopted a model-dependent method to determine the $H_0$ values for 5 different cosmological models by utilizing the latest  HII galaxy data. In this section, we would like to use the model-independent GP method to check the correctness of the $H_0$ values from the model-dependent method.

We use the publicly available package GaPP (Gaussian processes in python) to implement our reconstruction \cite{41}. The GP is a generalization of a Gaussian distribution, which is a distribution of a random variable, and exhibits a distribution over functions. The GP can reconstruct directly a function from the observed data without assuming a specific parameterization for the underlying function. At each reconstruction point $z$, the reconstructed function $f(z)$ is a Gaussian distribution with a mean value and Gaussian error. The key of the GP is a covariance function $k(z,\tilde{z})$ depending on two hyper-parameters $l$ and $\sigma_f$, which characterize the coherent scale of the correlation in $x$-direction and typical change in $y$-direction, respectively. In \cite{42}, the authors has verified the Mat\'{e}rn ($\nu=9/2$) covariance function is a better choice to carry out the reconstruction than the usual squared exponential covariance function. Therefore, we adopt the Mat\'{e}rn ($\nu=9/2$) covariance function to exhibit the first GP reconstruction using the HII galaxy data in the literature:
\begin{equation}
k(x,\tilde{x})=\sigma_f^2 \mathrm{exp}(-\frac{3|z-\tilde{z}|}{l})\times[1+\frac{3|z-\tilde{z}|}{l}+\frac{27(z-\tilde{z})^2}{7l^2}+\frac{18|z-\tilde{z}|^3}{7l^3}+\frac{27(z-\tilde{z})^4}{35l^2}]. \label{11}
\end{equation}
As the previous works \cite{41,43,44}, using Eq. (\ref{3}) and the expression of the normalized comoving distance $D(z)=H_0(1+z)^{-1}d_L(z)$, we transform the theoretical distance modulus $\mu_{th}$ to $D(z)$ in the following manner
\begin{equation}
D(z)=\frac{H_0}{1+z}10^{\frac{\mu_{th}-25}{5}}, \label{12}
\end{equation}
and the dark energy EoS can be expressed as
\begin{equation}
\omega(z)=\frac{2(1+z)(1+\Omega_{k0})D''-[(1+z)^2\Omega_{k0}D'^2-3(1+\Omega_{k0}D^2)+2(1+z)\Omega_{k0}DD']D'}{3D'\{(1+z)^2[\Omega_{k0}+(1+z)\Omega_{m0}]D'^2-(1+\Omega_{k0}D^2)\}}, \label{2}
\end{equation}
where the prime represents the derivative with respect to the redshift $z$. As noted in \cite{41}, we set the initial conditions $D(z=0)=0$ and $D'(z=0)=1$ throughout the reconstruction processes. Note that the $H_0$ value affects obviously the reconstruction results by affecting the transformed $D(z)$.
Additionally, we have assumed $\Omega_{k0}=0$ and $\Omega_{k0}=0.308\pm0.012$ \cite{45} in our GP reconstructions.  

In Fig. \ref{f3}, we consider the effects of 3 different $H_0$ values on the reconstructions of $D(z), D'(z)$, $D''(z)$ and dark energy EoS, i.e., R16's result $H_0=66.93\pm0.62$ km s$^{-1}$ Mpc$^{-1}$, P16's result $H_0=73.24\pm1.74$ km s$^{-1}$ Mpc$^{-1}$ and the prediction of the $\Lambda$CDM model $H_0=76.12^{+3.47}_{-3.44}$ km s$^{-1}$ Mpc$^{-1}$ using the HII galaxy data. From upper panels of Fig. \ref{f3}, one can easily find that, when $H_0=76.12^{+3.47}_{-3.44}$ km s$^{-1}$ Mpc$^{-1}$, the reconstructions of $D(z), D'(z)$ and $D''(z)$ are better than the other two cases. In the lower left and medium panels of Fig. \ref{f3}, when $H_0=66.93\pm0.62$ km s$^{-1}$ Mpc$^{-1}$ and $73.24\pm1.74$ km s$^{-1}$ km s$^{-1}$ Mpc$^{-1}$, we find that the underlying true model is consistent with the $\Lambda$CDM model in the low-redshift range at $2\sigma$ confidence level. However, when $H_0=76.12^{+3.47}_{-3.44}$ km s$^{-1}$ Mpc$^{-1}$, this occurs in the low-redshift range at $1\sigma$ confidence level. Hence, we conclude that the GP reconstructions prefer a higher $H_0$ value than the measurements in previous works (e.g., R16 and P16), and that the model-independent method have verified the correctness of $H_0$ values obtained by model-dependent method. It is worth noting that, because the current HII galaxy data points are mainly located at low redshifts (low-$z$ and GEHR samples) and there is a lack of medium and high redshifts data, we can not provide a accurate constraint on $D(z), D'(z)$, $D''(z)$ and the dark energy EoS.          

\section{Discussions and conclusions}
Precise measurements of $H_0$ is one of the most important and intriguing tasks. The recent local measurement implemented by Riess et al. exhibits a high tension with the Planck's result from CMB anisotropy data at $3.4\sigma$ confidence level. Our motivation is to use the latest HII galaxy measurements to determine the value of $H_0$.          

We explore the value of $H_0$ by using a combination of model-dependent and model-independent method. First of all, we constrain five cosmological models by using the newest compilation of HII galaxy measurements and obtain the corresponding values of $H_0$. We find that the $H_0$ values for these five cosmological models are all consistent with the R16's local measurement at $1\sigma$ confidence level, and that these five models prefer a higher best-fit $H_0$ value than R16's result (see Table \ref{t2} and Fig. \ref{f1}). In light of this, to check the correctness of $H_0$ values obtained by model-dependent method, we firstly implement the GP reconstructions using the HII galaxy data in the literature. We find that, when $H_0=76.12^{+3.47}_{-3.44}$ km s$^{-1}$ Mpc$^{-1}$, the reconstructions of $D(z), D'(z)$ and $D''(z)$ are better than the other two cases (e.g. R16 and P16), and the reconstructed dark energy EoS is more consistent with the $\Lambda$CDM model in the low-redshift range at $1\sigma$ confidence level than the other two cases (see Fig. \ref{f3}). Hence, one can conclude that the GP reconstructions prefer a higher $H_0$ value than R16's and P16's results, and that the model-independent method have verified the correctness of $H_0$ values obtained by model-dependent method.

In \cite{C}, using 69 nearby HII galaxies and 23 GEHR in 9 galaxies, the authors have obtained a value $H_0=74.3\pm3.1(\mathrm{statistical})\pm2.9(\mathrm{systematic})$, which is also consistent with our results at $1\sigma$ confidence level. However, unlike their results, we have obtained higher best-fit $H_0$ values by utilizing a combination of model-dependent and model-independent method. 

The tendency of high $H_0$ values may be attributed to different data systematics, a lack of medium and high redshift data or underlying new physics. In addition, from the point of view of the HII galaxies Hubble diagram, we find that there exists a very high degeneracy among these five models using only the latest HII galaxy measurements (see Fig. \ref{f2}). In the future, we expect more and more high-precision data can provide more useful information for us and help us distinguish different cosmological models better.

\section{acknowledgements}
 We are thankful to Professors S. D. Odintsov and Bharat Ratra for beneficial discussions on cosmology. The author Deng Wang warmly thanks Prof. Jing-Ling Chen for talks on Bell inequality. This work is supported in part by the National Science Foundation of China.

\end{document}